\documentclass[a4paper,12pt]{article}
\usepackage[left=2cm,right=2cm]{geometry}
\usepackage{graphics}
\begin{document}
\title{Three-body hadronic structure of low-lying $1/2^+$ $\Sigma$ and $\Lambda$ resonances}
\author{A. Mart\'inez Torres,  K. P. Khemchandani and E. Oset}                    

\date{}
\maketitle
\begin{center}
Departamento de F\' isica Te\' orica and IFIC, Centro Mixto Universidad de Valencia-CSIC, 
Institutos de Investigaci\'on de Paterna, Aptd. 22085, 46071 Valencia, Spain.
\end{center}
\abstract{We discuss the dynamical generation of some low-lying $1/2^+$ $\Sigma$'s and $\Lambda$'s
in two-meson one-baryon systems. These systems have been constructed by adding a pion in $S$-wave to the $\bar{K} N$ pair and its coupled channels, where the $1/2^-$ $\Lambda$(1405)-resonance gets dynamically generated. We solve Faddeev equations in the coupled-channel approach to calculate the $T$-matrix for these systems as a function of the total energy and the invariant mass of one of the meson-baryon pairs. This squared $T$-matrix shows peaks at the energies very close to the masses of the strangeness -1,  $1/2^+$ resonances listed in the particle data book.

\section{Introduction}
\label{intro}
Several claims of excitation of a narrow resonance at $\sim$ 1540 MeV were made in the $p K_s$ final state in different experiments \cite{many}, which were performed to confirm the existence of the pentaquark state, $\Theta^+$  \cite{Nakano}. The $p K_s$ system could possess a total strangeness +1 or -1 and, hence, these states could very well correspond to, for example, a $\Sigma^*$ \cite{zhao,close}. Actually, the existing $\Sigma$ and $\Lambda$ resonances \cite{pdg}, in this energy region, are not well understood. The poor status of these low-lying, $S=-1$ states is evident from the following facts: a) The spin-parity assignment for many of these states is unknown, e.g., for $\Sigma$(1480), $\Sigma$(1560), etc. b) the partial-wave analyses and production experiments have been often kept separately in the particle data group listings, {\it e.g.}, for $\Sigma$(1620), $\Sigma$(1670), c) other times, {\it e.g.}, in case of $\Lambda$ (1600), it is stated that existence of two resonances, in this energy region, is quite possible \cite{pdg}.

There are hints for some of them, like the $\Lambda(1600)$ and the $\Sigma(1660)$, to decay to three-body final states like $\pi^0\pi^0\Lambda$ and $\pi^0\pi^0\Sigma$, etc., \cite{prakhov,prakhov2}. This indicates that the information of similar three-body contributions to the wave functions of the states, in this energy region, could be important and that this, so far unknown, information could lead to a better understanding of the low-lying $\Sigma$ and $\Lambda$ resonances. The $\pi$-$\Sigma$ interaction has been
studied extensively in chiral unitary dynamics, where the $\Lambda(1405)$ gets dynamically generated (with a two pole structure \cite{jido}). Is it possible that adding another pion to this strongly attractive system could give rise to a three-body bound state? We have carried out a study of such systems for total spin-parity ($J^P$) = $1/2^+$ and, indeed, find evidence for four $\Sigma$ and two $\Lambda$ known resonances \cite{pdg} to have a three body, i.e., two-meson one-baryon, structure, which we discuss in the following text. The chiral dynamics has been used earlier in the context of the three-nucleon problems, {\it e.g.}, in \cite{epelbaum}. This is the first time when the chiral dynamics is applied to solve the Faddeev equations for two-meson one-baryon systems.

\section{Formalism}
\label{sec:1}
We solve Faddeev equations in the coupled channel approach. These coupled systems have been constructed by pairing up all the possible pseudoscalar mesons and $1/2^+$ baryons, which couple to strangeness =-1 and by adding a pion to the pair finally. We end up with twenty-two coupled channels for a fixed total charge \cite{mko}.
In order to generate dynamically the $\Lambda(1405)\,S_{01}$ ($J^P=1/2^-$) in the meson-baryon sub-systems, all the interactions have been written in the $S$-wave, which implies that the total $J^P$ of the three body system is $1/2^+$. 

The solution of the Faddeev equations,
\begin{equation}\label{ft}
T = T^1 + T^2 + T^3,
\end{equation}
where the $T^i$ partitions are written in terms of two body t-matrices ($t^i$) and three body propagators
($g = [E - H]^{-1}$) as
\begin{equation}\label{fp}
T^{i}=t^i + t^i g \Big[ T^{j} + T^{k} \Big] \,\,\,\, {\rm (i \neq j, \,\,i,j \neq k =1,2,3)},
\end{equation}
requires off-shell two body $t$-matrices as an input. However, we found that the off-shell contributions of the two body $t$-matrices give rise to three-body forces which, when summed up for different diagrams, cancel the one arising directly from the chiral Lagrangian in the $SU(3)$ limit and for the case of small momentum transfer of the baryon \cite{mko}. In the realistic case, this sum was found to be only 5$\%$ of the total on-shell contribution \cite{mko}. Thus, it is reasonably accurate to study the problem by solving the Faddeev equations with the on-shell two-body $t$-matrices and by neglecting the three-body forces.

In this way, a term at second order in $t$ in the Faddeev partitions (fig. \ref{figtgt}), for instance, 
\begin{equation}\label{t1gt3}
t^1 g^{13} t^3,
\end{equation}
is written as a product of the on-shell two body $t$-matrices $t^1$ and $t^3$
\begin{figure}[ht]
\begin{center}
\resizebox{0.25\textwidth}{!}{%
\includegraphics{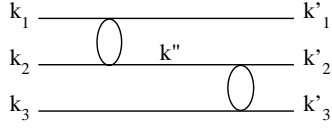}
}
\caption{\textit{The diagrammatic representation of the $t^1gt^3$ term.}}\label{figtgt}
\end{center}
\end{figure}
and the propagator $g^{13}$ given as
\begin{eqnarray}
g^{13} = \frac{1}{2E_2}\frac{1}{\sqrt{s}-E_1
(\vec{k}^\prime_1)-E_2(\vec{k}^\prime_1+\vec{k}_3)-E_3(\vec{k}_3)+i\epsilon}\nonumber.
\end{eqnarray}

Adding another interaction to the diagram in fig. \ref{figtgt} (see fig. \ref{tgtgt}), the expression $t^1g^{13}t^3$
\begin{figure}[ht]
\begin{center}
\resizebox{0.25\textwidth}{!}{%
\includegraphics{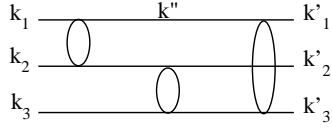}
}
\caption{\textit{The diagrammatic representation of $t^2g^{21}t^1g^{13}t^3$.}}\label{tgtgt}
\end{center}
\end{figure}
gets extended to\\
$t^2g^{21}t^1g^{13}t^3$, which can be written explicitly as
\begin{eqnarray}
&&t^2(s_{13})\Bigg[\int \frac{d\vec{k}^ {\prime\prime}}{(2\pi)^3} \frac{1}{2E_1(\vec{k}^{\prime\prime})} \frac{1}{2E_2(\vec{k}^{\prime\prime}+\vec{k}_3)} \frac{2 M_3}{2 E_3(\vec{k}^{\prime\prime}+ \vec{k}^\prime_2)}\nonumber\\
&&\times\frac{t^1(s_{23}(\vec{k}^{\prime\prime}))}{\sqrt{s}-E_1
(\vec{k}^{\prime\prime})-E_2(\vec{k}^{\prime}_2)-E_3(\vec{k}^{\prime\prime}+\vec{k}^{\prime}_2)+i\epsilon}\nonumber\\
&&\times\frac{1}{\sqrt{s}-E_1(\vec{k}^{\prime\prime})-E_2(\vec{k}^{\prime\prime}+\vec{k}_3)-E_3(\vec{k}_3)+i\epsilon}\Bigg]t^3(s_{12}),\nonumber
\end{eqnarray}
where $s_{13} = (P - K^{\prime}_2)^2$, $s_{12} = (P - K_3)^2$, $s_{23}(\vec{k}^{\prime\prime}) = (P - K^{\prime\prime})^2$ with $P, K_3, K^{\prime}_2$ and $K^{\prime\prime}$ representing the four momenta corresponding to the $\vec{P} = 0, \vec{k}_3, \vec{k}^{\prime}_2$ and $\vec{k^{\prime\prime}}$ respectively.
Our aim is to extract $t^1g^{13}t^3$ out of the integral, which could simplify the calculations. The $t^2(s_{13})$ and $t^3(s_{12})$, in the equation above , depend on on-shell variables and can be factorized out of the loop integral but not $t^1$ and $g^{13}$ \cite{mko} . This can be done if we re-arrange the loop integral as
\begin{eqnarray}
\nonumber
&&t^2(s_{13}) \int \frac{d\vec{k}^ {\prime\prime}}{(2\pi)^3}\frac{1}{2E_1(\vec{k}^{\prime\prime})} \frac{2 M_3}{2 E_3(\vec{k}^{\prime\prime}+\vec{k}^\prime_2))}\frac{1}{\sqrt{s}-E_1(\vec{k}^{\prime\prime})-E_2(\vec{k}^{\prime}_2)-E_3(\vec{k}^{\prime\prime}+\vec{k}^{\prime}_2)+i\epsilon}\nonumber\\
&&\times\Bigg\{t^1(s_{23}(\vec{k}^{\prime\prime}))\frac{\displaystyle\frac{1}{2E_2(\vec{k}^{\prime\prime}+\vec{k}_3)}\frac{1}{\sqrt{s}-E_1(\vec{k}^{\prime\prime})-E_2(\vec{k}^{\prime\prime} + \vec{k}_3) - E_3(\vec{k}_3)+i\epsilon}}
{\displaystyle\frac{1}{2E_2(\vec{k}^{\prime}_1+\vec{k}_3)}\frac{1}{\sqrt{s}-E_1(\vec{k}^{\prime}_1) - E_2(\vec{k}^{\prime}_1+\vec{k}_3) - E_3(\vec{k}_3)+i\epsilon}}
\times[t^1(s_{23})]^{-1} \Bigg\}\nonumber\\
&&\times\Bigg(t^1(s_{23}) \frac{1}{2E_2(\vec{k}^{\prime}_1+\vec{k}_3)}\frac{1}{\sqrt{s}-E_1(\vec{k}^{\prime}_1) - E_2(\vec{k}^{\prime}_1+\vec{k}_3) - E_3(\vec{k}_3)+i\epsilon}
t^3(s_{12})\Bigg).\nonumber
\end{eqnarray}
The expression in the circular brackets, in the above equation, corresponds to $t^1g^{13}t^3$ term which can be now extracted out of the loop integral since it depends on the on-shell variables. The off-shell dependence of $t^1$ and $g^{13}$ has been included in the terms enclosed in the curly brackets, which we define as $F^{213}$ and name as off-shell factor (which should not be confused with the off-shell ``form-factor'', often used in the literature. The off-shell factor, here, is just the nomenclature for a mathematical expression which retains the off-shellness of the terms which are extracted out of the integral). The integral is now left with two propagators and the off-shell factor $F^{213}$, which together we re-express and define as a loop function, $G^{213}$,
\begin{eqnarray}
G^{213} = \int \frac{d\vec{k}^{\prime\prime}}{(2\pi)^3}\frac{1}{2E_1(\vec{k}^{\prime\prime})}\frac{2M_3}{2E_3(\vec{k}^{\prime\prime})}\frac{F^{213}(\sqrt{s},\vec{k}^{\prime\prime})}{\sqrt{s_{13}} - E_1(\vec{k}^{\prime\prime}) - E_3(\vec{k}^{\prime\prime})+i\epsilon}\nonumber
\end{eqnarray}
The diagram in fig. \ref{tgtgt} can, hence, be re-written as $t^2G^{213}$ $t^1g^{13}t^3$.
The formalism has been developed following the above procedure, i.e., by replacing $g$ by $G$, everytime a new interaction is added. This leads to another form of the Faddeev partitions (eq. (\ref{fp})), which, after removing the terms corresponding to the disconnected diagrams and by denoting the rest of the equation as $T_R$, can be re-written as 
\begin{equation}
T_R^{ij}=t^i g^{ij} t^j + t^i G^{ijk} T_R^{jk} + t^i G^{iji} T_R^{ji}, \,\,\,\, {\rm i \neq j, j \neq k =1,2,3}.
\end{equation}

The $T_R^{ij}$ can be related to the Faddeev partitions (eq. (\ref{fp})) as $T^i = t^i + T_R^{ij} + T_R^{ik}$, hence, giving six coupled equations instead of three (eq. (\ref{fp})). These $T_R^{ij}$ partitions correspond to the sum of all the possible diagrams with the last two interactions written in terms of $t^i$ and $t^j$. This re-grouping of diagrams is done for the sake of convenience due to the different forms of the $G^{ijk}$ functions. We define $T_R$ as 
\begin{equation}\label{tr}
T_R = \sum_{i \neq j = 1}^3 T_R^{ij},
\end{equation}
which can be related to the sum of the Faddeev partitions (eq. (\ref{ft})) as 
\begin{equation}
T = t^1 + t^2 + t^3 + T_R.
\end{equation}

\section{Results and discussion}\label{sec:2}
We shall construct the three-body $T_R$-matrices using the isospin symmetry, for which we must take an average mass for the isospin multiplets $\pi$ $(\pi^+,\,\pi^0,\,\pi^-)$, $\bar{K}$ $(\bar{K}^0,\,K^-)$, $K$ $(K^+,\, K^0)$, $N$ $(p,\,n)$, $\Sigma$ $(\Sigma^+,\,\Sigma^0,\,\Sigma^-)$ and $\Xi$ $(\Xi^0,\,\Xi^-)$. In order to identify the nature of the resulting states, we project the $T_R$-matrix on the isospin base. One appropriate base is the one where the states are classified by the total isospin of the three particles, $I$, and the total isospin of the two mesons, $I_\pi$ in the case of two pions. Using the phase convention $\mid \pi^+\rangle=-\mid 1,1\rangle$, $\mid K^-\rangle=-\mid 1/2,-1/2\rangle$, $\mid \Sigma^+\rangle=-\mid 1,1\rangle$ and $\mid \Xi^-\rangle=-\mid 1/2,-1/2\rangle$ we have, for example, for the the $\pi\,\pi\,\Sigma$ channel
\begin{eqnarray}
\nonumber
\mid \pi^0\,&\pi^0&\,\Sigma^0\rangle =\mid 1, 0\rangle_{\pi}\,\otimes\mid 1, 0\rangle_{\pi}\,\otimes\mid 1, 0\rangle_{\Sigma} \nonumber\\
&=&\left\{\sqrt{\frac{2}{3}}\mid 2, 0\rangle-\sqrt{\frac{1}{3}}\mid 0, 0\rangle\right\}_{\pi\pi}\,\otimes
 \mid 1, 0\rangle_{\Sigma}\nonumber\\
&=&\sqrt{\frac{2}{5}}\mid I=3,I_\pi=2\rangle-\frac{2}{\sqrt{15}}\mid I=1,I_\pi=2\rangle-\nonumber\\
&&-\sqrt{\frac{1}{3}}\mid I=1,I_\pi=0\rangle,\nonumber
\end{eqnarray}
where, $I$ and $I_{\pi}$ denote the total isospin of the three body system and that of the two pion system, respectively. Similarly, we write the other states in the isospin base. 

\begin{figure}[hb]
\resizebox{1\textwidth}{!}{%
  \includegraphics{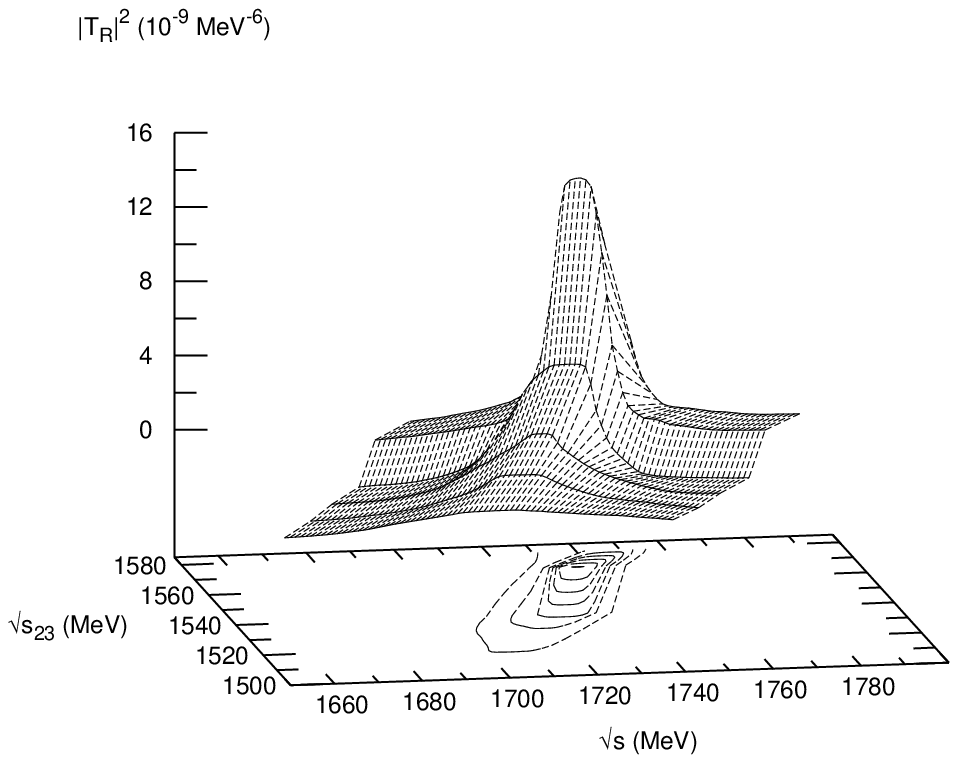}
}
\caption{The $\Lambda$(1810) resonance depicted in the squared amplitude for $\pi \pi \Lambda$ in the $\mid I, I_{\pi}\rangle$ = $\mid 0, 0\rangle$ configuration.}
\label{fig:1}       
\end{figure}

After projecting the $T_R$-matrix (eq. (\ref{tr})) on the isospin base, we square them and plot them as a function of the total energy ($\sqrt{s}$) of the two meson one baryon system and the invariant mass of particles $2$ and $3$ ($\sqrt{s_{23}}$). However, this choice is arbitrary, we could as well plot the squared amplitude, for example,  as a function of $\sqrt{s_{12}}$ and $\sqrt{s_{13}}$.

In fig. \ref{fig:1}, we show our results for $\pi \pi \Lambda$ in the $\mid I, I_{\pi}\rangle$ = $\mid 0, 0\rangle$ case, where a peak at 1740 MeV is clearly seen. The full width at half maximum of this peak is $\sim$ 20 MeV. We identify this resonant structure with the $\Lambda$(1810) in the particle data book \cite{pdg}.  It should be noted that, even though the status of this resonance in \cite{pdg} is three-star, the associated pole positions vary from about 1750 MeV to 1850 MeV and the corresponding widths from 50-250 MeV. 

We find evidence for another resonance in the isospin zero sector in the $\Lambda$(1600) region. Apart from this, we find evidence for four low-lying $\Sigma$ resonances: 1) for the $\Sigma$(1560), thus predicting $J^P =1/2^+$ for it, 2) the $\Sigma$(1620), for which the partial-wave analyses and experimental findings are kept separately in the $PDG$, 3) the well-established $\Sigma$(1660) and 4) the one-star $\Sigma$(1770). All these results are summarized in the table 1, along with the states listed by the Particle Data Group.

\begin{table}[h]
\begin{center}
\caption{A comparison of the resonances found in this work with the states in $PDG$.}
\label{tab:1}       
\vspace{0.5 cm}
\begin{tabular}{lccc}
\hline\noalign{\smallskip}
& $\Gamma$ ($PDG$) & Peak position  & $\Gamma$ (this work)\\
& (MeV) & (this work, MeV) & (MeV)\\
\noalign{\smallskip}\hline\noalign{\smallskip}
\multicolumn{4}{l}{Isospin = 1} \\
\noalign{\smallskip}\hline\noalign{\smallskip}
$\Sigma(1560)$&10 - 100&1590&70\\
$\Sigma(1620)$&10 - 100&1630&39\\
$\Sigma(1660)$&40 - 200&1656&30\\
$\Sigma(1770)$&60 - 100&1790&24\\
\noalign{\smallskip}\hline
\multicolumn{4}{l}{Isospin = 0} \\
\noalign{\smallskip}\hline
$\Lambda(1600)$&50 - 250&1568 &60\\
$\Lambda(1810)$&50 - 250&1740&20\\
\noalign{\smallskip}\hline

\end{tabular}
 \end{center}
\end{table}


%

\begin{thebibliography}{}
%
%
\bibitem{many}DIANA Collaboration (V.~V.~Barmin {\it et al.})  ,
  Phys.\ Atom.\ Nucl.\  {\bf 66}, 1715 (2003); HERMES Collaboration (A.~Airapetian {\it et al.})  ,
  Phys.\ Lett.\  B {\bf 585}, 213 (2004);  SVD Collaboration (A.~Aleev {\it et al.})  ,
  Phys.\ Atom.\ Nucl.\  {\bf 68}, 974 (2005);  ZEUS Collaboration (S.~Chekanov {\it et al.})  ,
  Phys.\ Lett.\  B {\bf 591}, 7 (2004);
  COSY-TOF Collaboration (M.~Abdel-Bary {\it et al.})  , Phys.\ Lett.\  B {\bf 595}, 127 (2004), etc.

\bibitem{Nakano}
T.~Nakano {\it et al.}, Phys.\ Rev.\ Lett.\  {\bf 91},  (2003) 012002.

\bibitem{zhao}
  Q.~Zhao and F.~E.~Close,
  J.\ Phys.\ G {\bf 31}, L1 (2005).

\bibitem{close}
  F.~E.~Close and Q.~Zhao,
Phys.\ Lett.\  B {\bf 590}, 176 (2004).

\bibitem{pdg} W.-M. Yao {\it et al.}, J. Phys. G \textbf{33}, 1 (2006).

\bibitem{prakhov}
  S.~Prakhov {\it et al.},
  Phys.\ Rev.\  C {\bf 70}, 034605 (2004).


\bibitem{prakhov2}
  S.~Prakhov {\it et al.},
  Phys.\ Rev.\  C {\bf 69} (2004) 042202.

  S.~Prakhov {\it et al.}  [Crystall Ball Collaboration],
  Phys.\ Rev.\  C {\bf 70}, 034605 (2004).


\bibitem{jido}
  D.~Jido {\it et. al.},
  Nucl.\ Phys.\  A {\bf 725}, 181 (2003).

\bibitem{epelbaum}
  E.~Epelbaum {\it et. al.},
  Phys.\ Rev.\ Lett.\  {\bf 86}, 4787 (2001).

\bibitem{mko}
  A.~Martinez Torres, K.~P.~Khemchandani and E.~Oset,
  Phys.\ Rev.\  C {\bf 77}, 042203 (2008) .
\end{thebibliography}
%

\end{document}